\documentclass[prb,preprint]{revtex4}
\usepackage{amsmath}

\newcommand{\A}{\textsf{A}}
\newcommand{\T}{\textsf{T}}
\newcommand{\bA}{\bfseries{\sffamily{A}}}
\newcommand{\bT}{\bfseries\sffamily{T}}
\newcommand{\lb}{{<}}
\newcommand{\rb}{{>}}

\begin{document}

\title{The EPR paradox, Bell's inequality, and the question of locality}

\author{Guy Blaylock}
\email{blaylock@physics.umass.edu} 
\affiliation{University of Massachusetts, Department of 
Physics, Amherst, MA 01003}

\begin{abstract}
Most physicists agree that the Einstein-Podolsky-Rosen-Bell paradox exemplifies much of the strange behavior of quantum mechanics, but argument persists about what assumptions underlie the paradox. To clarify what the debate is about, we employ a simple and well-known thought experiment involving two correlated photons to help us focus on the logical assumptions needed to construct the EPR and Bell arguments. The view presented in this paper is that the minimal assumptions behind Bell's inequality are locality and counterfactual definiteness, but not scientific realism, determinism, or hidden variables, as is often suggested. We further examine the resulting constraints on physical theory with an illustration from the many-worlds interpretation of quantum mechanics -- an interpretation that we argue is deterministic, local, and realist, but that nonetheless violates the Bell inequality.\end{abstract}

\maketitle

\section{INTRODUCTION}

\textit{``Like all authors of noncommissioned reviews, [this author] thinks that he can restate the position with such clarity and simplicity that all previous discussions will be eclipsed."} -- J.~S.\ Bell.\cite{Bell66}

In 1935, Einstein, Podolsky, and Rosen [EPR] argued that quantum mechanics cannot be a complete theory of the physical world.\cite{Einstein35} They based their argument on an analysis of two separate but correlated particles, for which a characteristic of one particle can be determined by measuring the other particle.\cite{EPRexamples} Assuming the measurement of the second particle does not affect the first particle (no nonlocal effects), EPR claimed that because the characteristic of the first particle can be precisely determined without directly measuring it (that is, by measuring only the correlated particle), that characteristic must have a well-defined existence independent of the observer. In their words, it must be an ``element of reality'' and therefore ought to be described by any theory that claims to be complete. Because quantum mechanics does not in general describe the precise values of physical characteristics --- it only makes statistical statements describing the likelihood of different values --- quantum theory must therefore be incomplete. A properly complete theory, they implied, must contain extra ``hidden variables'' that would describe the particle characteristic precisely. 

In constructing this argument, EPR adhered to the common sense principle of ``scientific realism'' (that measurable characteristics of a physical system exist and are well defined, independent of any outside influence or observation). The ``elements of reality'' represent their view of that underlying realism. Furthermore, assuming the ``elements'' evolve in a predictable fashion, the principle of determinism (that complete knowledge of the current state of a physical system can be used to determine the future state of the system) is also valid. In this way, the EPR ``elements'' allow precise predictions of measurement results and offer a potential alternative to the statistical nature of orthodox quantum mechanics. For decades following the EPR publication debate raged over the validity of these principles and what kind of predictions should be expected from a physical theory, without providing any clear alternative to quantum mechanics.

In 1964, John Bell published a related thought experiment\cite{Bell65} that distinguished between quantum mechanics and all local hidden-variable theories. In John Bell's version of the EPR thought experiment, quantum mechanics and local hidden-variable theories predicted statistically different experimental results. What had previously been a philosophical debate suddenly had testable consequences. Bell's experiment has since been performed in the laboratory,\cite{experiments} verifying the predictions of quantum mechanics and dramatically contradicting Bell's predictions for local hidden-variable theories.\cite{limitations}

Since then, debate has focused on the foundation of Bell's predictions. Bell's original proof concentrates on theories with hidden variables (that is, variables that ``determine precisely the results of individual measurements,"\cite{Bell66}) which are inherently realist.\cite{HVrealism} Some authors state that the derivation applies specifically to deterministic hidden variables.\cite{HVdeterminism} Other authors suggest that Bell's predictions rest more generally on local realism\cite{localrealists} (without being necessarily deterministic) or on locality alone.\cite{localists} Broadly speaking, there are two camps of thought on this issue: there are those who think that Bell's inequality can be derived purely from the assumption of locality (and the experimentally measured correlations in the EPR experiment), and those who think that Bell's theorem requires some other assumption as well.

In this paper we use a carefully broad definition of ``locality'' (see the following) and support the latter view. However, we argue that realism, determinism, and hidden variables do not provide the minimal necessary additional assumption, and that these assumptions have only incidental connection with the logic of the Bell derivation. Instead, following the comments of previous authors,\cite{CFDers} we claim that the minimal assumptions behind Bell's predictions are locality and ``counterfactual definiteness'' (that we can postulate a single definite result from a measurement even when the measurement is not performed), and we illustrate the subtle distinctions between these various assumptions with some examples.

There is some risk in trying to clarify such a contentious issue. Although the role of counterfactual definiteness in the Bell analysis was identified as early as 1971,\cite{Stapp71} it is not broadly accepted by the physics community, and there is much debate about the interpretation of the Bell experiment. In most cases, the abstract nature of the argument inhibits explanation to lay audiences and even confounds the discussion for professional physicists. To try to alleviate this confusion we focus on a simple concrete example to illustrate the logic in Bell's derivation and clarify the assumptions about locality, hidden variables, realism, determinism, and counterfactual definiteness. Hopefully, this approach can be used as the basis for explanation to a lay audience and to help clarify the concepts for professional scientists (as it has for this author). We conclude by briefly describing some popular interpretations of quantum mechanics (focusing especially on the many-worlds interpretation) as a way of further illustrating the various assumptions about an underlying physical reality.
	
\section{THE EPR THOUGHT EXPERIMENT}

In their paper,\cite{Einstein35} EPR explored the predictions for position and momentum measurements of correlated two-particle systems. In 1950, David Bohm reformulated the EPR experiment as measurements of spin 1/2 particles\cite{Bohm51} instead of the continuous range of results from position or momentum measurements. In our approach we consider measurements of photon polarizations, which follow the approach of many real-life EPR experiments and are closely analogous to Bohm's thought experiment. Photon polarizations are easier to measure than the orientation of spin 1/2 particles --- a polarizing filter\cite{polarizers} can be used instead of a Stern-Gerlach apparatus --- and are easier to explain to the philosopher on the street because they don't require the concept of spin. The photons pass through or are absorbed by a polarizing filter, and the probability of a photon with a certain (linear) polarization passing through an ideal polarizing filter at a certain orientation can be calculated using simple geometry.

We begin by imagining two photons that are traveling side-by-side in the same direction (but separated from each other), and which are prepared in the ``twin state'' polarization. In quantum mechanics the term ``twin state'' has a particular meaning, but our analysis of Bell's logic requires only an experimental definition. In practical terms, the ``twin state'' means that measurements of the two photons using identical polarizing filters always produce the same result. If the first photon passes through a polarizing filter, the second photon is guaranteed to pass through another filter oriented at the same angle. If the first photon is absorbed, the second photon will also be absorbed, no matter the orientation of the filters, as long as the two filters are pointing in the same direction. The results of the polarization measurements are perfectly correlated.\cite{twinstates}

In quantum mechanics the twin state is represented by an entangled isotropic state such as
\begin{equation}
\label{eq:1}
\psi=\frac{1}{\sqrt{2}}\left(|x\rangle_1|x\rangle_2+|y\rangle_1|y\rangle_2\right),
\end{equation}
where $|x\rangle_i$ represents a polarization state of the $i$th photon that is guaranteed to pass through a horizontal filter (and guaranteed to be absorbed by a vertical filter) and $|y\rangle_i$ represents a polarization state that is guaranteed to pass through a vertical filter (and guaranteed to be absorbed by a horizontal filter). In this form it is clear that if photon~1 registers as having vertical polarization, photon~2 also turns out to be vertical. In addition, we recognize that the twin state is rotationally symmetric by rotating the $x$ and $y$ axes by an arbitrary angle:
\begin{subequations}
\label{eq:2}
\begin{align}
\psi &= \frac{1}{\sqrt{2}}\left(|x\rangle_1|x\rangle_2+|y\rangle_1|y\rangle_2\right)\\
&= \frac{1}{\sqrt{2}}(\left[|x\rangle_1'\cos\theta-|y\rangle_1'\sin\theta\right]\left[|x\rangle_2'\cos\theta-|y\rangle_2'\sin\theta\right] \nonumber \\
&{} \quad +\left[|y\rangle_1'\cos\theta+|x\rangle_1'\sin\theta\right]\left[|y\rangle_2'\cos\theta+|x\rangle_2'\sin\theta\right])\\
&= \frac{1}{\sqrt{2}}\left(|x\rangle_1'|x\rangle_2'+|y\rangle_1'|y\rangle_2'\right),
\end{align}
\end{subequations}
where $x'$ and $y'$ are rotated counterclockwise from $x$ and $y$ by angle $\theta$. The fact that the form is the same in the rotated basis indicates that the state is rotationally symmetric. Although this representation is not crucial to our study, it satisfies our experimental definition of the twin state: no matter what direction we orient our filters, the two polarization measurements are always the same.

The original EPR argument applies to the twin state system as follows. Although initially we don't know the polarization of the first photon (it could either pass through or be absorbed by its filter), we can determine it by measuring the polarization of the second photon. Because by doing this measurement we determine the polarization of the first photon without disturbing it, EPR argue that the polarization of the first photon must therefore be a well-defined quantity, independent of its measurement. It follows that a complete theory should contain an element that specifies this polarization.

In stark contrast to this view, the orthodox interpretation of quantum mechanics claims that the polarizations of the photons are not well defined prior to measurement. Initially the two photons are in an entangled superposition of $|x\rangle$ and $|y\rangle$ states. Only when the measurement of one of the photons takes place do both photons collapse to well-defined polarizations. Quantum mechanics does not precisely specify what the polarizations are for a particular pair of photons before they are measured. Quantum mechanics does not even admit the definite polarization of either photon as a valid concept in the theory. Although the quantum mechanical twin state of 
Eq.~(1) specifies the correlation between photons --- whatever the polarization of one photon, the other is guaranteed to be the same --- it does not specify the individual polarization of either photon. Each photon is in an entangled superposition of ill-defined polarization. EPR argue therefore, following the principle of scientific realism, that quantum mechanical theory fails to specify an element of reality. As a physical theory, it is incomplete.

As is commonly recognized, and as EPR themselves admitted, this argument assumes that the measurement of the second photon does not in any way affect the state of the (spatially separated) first photon. By this reasoning, EPR assume locality: namely that interactions between objects are point-like (local) interactions whose immediate effects are confined to a single location, and that actions at one location do not immediately have any effect at a separate location. This assumption, when combined with the conclusions of special relativity, implies that no effect can travel faster than the speed of light in vacuum.\cite{locality}

This notion of locality is closely related to (but not exactly the same as) relativistic causality, which offers a very powerful logical constraint on theory. Relativistic causality states that no thing (neither object nor signal) can travel faster than the velocity of light in vacuum. (However, it does not rule out the possibility of some kind of information-free superluminal effect, such as quantum collapse.) If objects or information could be transmitted faster than the speed of light --- if for instance, a nonlocal effect could somehow be used to instantaneously transmit a signal --- special relativity implies\cite{FTL} that causality would be violated, and we would be in danger of a number of logical paradoxes, such as accidentally causing the death of our grandmother before we were born. 

Although EPR recognized the locality assumption in their argument, they believed that assumption was unassailable. Regarding the possibility that elements of reality could depend on non-local effects, they concluded:
\textit{``No reasonable definition of reality could be expected to permit this."}\cite{Einstein35}
	
\section{BELL'S INEQUALITY}

EPR did not dispute the accuracy of the statistical predictions of quantum mechanics. They argued that quantum mechanics should describe more than it does; it should precisely describe the outcomes of individual measurements. Decades later, John Bell focused on the statistical predictions themselves and asked how quantum mechanics could be experimentally distinguished from other possible theories of the world.

Bell's innovation was to consider polarization measurements for twin state photons that are not necessarily made at the same angle.\cite{Bellstates} Although such pairs of measurements are not perfectly correlated, we should be able to calculate the average statistical correlation. The astonishing result is that the correlation predicted by quantum mechanics is not the same as the correlation predicted by common sense. A difference exists at the statistical level, which can be unambiguously resolved by experiment.

To follow this line of thought, imagine making measurements of twin state photons with polarizing filters oriented at two different angles, $\theta_1$ for the first photon and $\theta_2$ for the second photon. What does quantum mechanics predict for the probability of getting the same measurement in both cases (either both photons transmitted or both absorbed)? According to quantum mechanics, whichever measurement is performed first\cite{refframe} collapses the entangled twin state superposition to a single polarization state that is identical for both photons. Thus, if photon~2 passes through a filter oriented at angle $\theta_2$, both photons are collapsed to a polarization state along that direction. If we wish to know what the probability is of getting the same measurement for photon~1, we need only figure out what the probability is for a photon with polarization along $\theta_2$ to pass through a filter oriented along $\theta_1$. This probability is easily calculated according to simple trigonometry. Any arbitrary linear polarization can be thought of as a superposition of polarization along the $\theta_1$ direction (which will pass through the filter) and perpendicular to the $\theta_1$ direction (which will be absorbed by the filter). For a wave polarized along the $\theta_2$ direction, the amplitude component along the $\theta_1$ direction is given by $\cos(\theta_2-\theta_1)$, and the probability for transmission, given by the wave amplitude squared, is $\cos^2(\theta_2-\theta_1)$. That is the prediction of quantum mechanics.

Now let us contrast this prediction with the prediction of common sense. This is where Bell comes in, and we apply his logic to our two-photon case.\cite{Herbert85} For measurements with filters along the same direction, we know that the probability of obtaining identical measurement results (both transmitted or both absorbed) is 100\%. A typical sequence of paired measurements, represented as either absorbed (\A) or transmitted (\T), might look like the following:

\begin{center}
\begin{tabular}{|l|c|c|c|c|c|c|c|c|c|c|c|c|c|c|c|c|c|c|c|c|c|c|c|c|}
\hline
Filter 1&\A&\T&\T&\A&\A&\T&\A&\T&\A&\T&\T&\T&\A&\T&\A&\A&\A&\A&\T&\A&\T&\T&\A&\T\\
\hline
Filter 2&\A&\T&\T&\A&\A&\T&\A&\T&\A&\T&\T&\T&\A&\T&\A&\A&\A&\A&\T&\A&\T&\T&\A&\T\\
\hline
\end{tabular}
\end{center}
\vspace{6pt}
We always see both photons absorbed or both transmitted, never one absorbed and the other transmitted. Similarly, we know that paired measurements with filters oriented perpendicular to each other will always give opposite results; whenever one photon passes through its filter, the other photon is absorbed.

Furthermore, it is reasonable to assume that filter measurements differing by intermediate angles (between $0^\circ$ and $90^\circ$) will yield intermediate correlations, and that a continuous range of angles gives rise to a continuous range of correlation. For example, imagine some angle between $0^\circ$ and $90^\circ$, call it $\alpha$, for which the mismatch between measurements is 25\% on average (the coincidence rate is 75\%). Thus, for filter~1 oriented at $\theta_1=0$ and filter~2 oriented at angle $\theta_2=\alpha$, typical sequences might be

\begin{center}
\begin{tabular}{|l|c|c|c|c|c|c|c|c|c|c|c|c|c|c|c|c|c|c|c|c|c|c|c|c|}
\hline
Filter 1&\A&\T&\T&\A&\A&\T&\A&\T&\A&\T&\T&\T&\A&\T&\A&\A&\A&\A&\T&\A&\T&\T&\A&\T\\
\hline
Filter 2&\A&\T&\T&\bT&\A&\T&\A&\T&\A&\bA&\T&\bA&\A&\T&\bT&\A&\bT&\A&\T&\A&\T&\T&\bT&\T\\
\hline
\end{tabular}
\end{center}
\vspace{6pt}
where boldface entries highlight where there is disagreement between the two sequences. On average, we expect 25\% of the cases will disagree, and in our illustration, six out of 24 measurements do so. The same average coincidence rate should be present if $\theta_1=-\alpha$ and $\theta_2=0$ (which corresponds to tilting one's head by angle $\alpha$) so that typical sequences in this case might be:
\begin{center}
\begin{tabular}{|l|c|c|c|c|c|c|c|c|c|c|c|c|c|c|c|c|c|c|c|c|c|c|c|c|}
\hline
Filter 1&\A&\bA&\bA&\A&\A&\T&\bT&\T&\A&\T&\T&\T&\A&\T&\A&\bT&\bT&\A&\T&\A&\T&\bA&\bT&\T\\
\hline
Filter 2&\A&\T&\T&\A&\A&\T&\A&\T&\A&\T&\T&\T&\A&\T&\A&\A&\A&\A&\T&\A&\T&\T&\A&\T\\
\hline
\end{tabular}
\end{center}

Now consider how the sequences should compare if $\theta_1=-\alpha$ and $\theta_2=\alpha$:
\begin{center}
\begin{tabular}{|l|c|c|c|c|c|c|c|c|c|c|c|c|c|c|c|c|c|c|c|c|c|c|c|c|}
\hline
Filter 1&\A&\A&\A&\A&\A&\T&\T&\T&\A&\T&\T&\T&\A&\T&\A&\T&\bT&\A&\T&\A&\T&\A&\bT&\T\\
\hline
Filter 2&\A&\T&\T&\T&\A&\T&\A&\T&\A&\A&\T&\A&\A&\T&\T&\A&\bT&\A&\T&\A&\T&\T&\bT&\T\\
\hline
\end{tabular}
\end{center}
where we have just copied the previous sequences of measurements from $\theta_1=-\alpha$ and $\theta_2=\alpha$ to represent typical sequences of measurements (though we do not always measure exactly the same sequences, just sequences consistent with the same statistical coincidence rates). Because the measurements of the first photon at angle $-\alpha$ disagree with measurements at angle zero by 25\%, and because the measurements of the second photon at angle $\alpha$ also disagree with measurements at angle zero by 25\%, we would expect that measurements of photon~1 at $-\alpha$ and measurements of photon~2 at $\alpha$ would disagree with each other by an average of at most 50\%. The disagreement would be a maximum of 50\% because there are some cases where both measurements disagree with the result at angle zero (highlighted in boldface) and therefore agree with each other. Thus, according to this logic, the average mismatch between measurements with $\theta_1=-\alpha$ and $\theta_2=\alpha$ should be less than or equal to 50\%. This constraint is an example of Bell's inequality.\cite{inequality}

The startling consequence of this logic is that it conflicts with quantum mechanics. The ``common sense'' prediction states that the mismatch rate should be at most 50\%, while quantum mechanics predicts (and experiment confirms) that the mismatch rate in this case is 75\%. If we recall that the coincidence rate predicted by quantum mechanics is $\cos^2(\theta_2-\theta_1)$, we recognize that the mismatch rate should be 25\% when $\alpha=\theta_2-\theta_1=30^\circ $ (that is, $\cos^2(30^\circ)=0.75)$. When $\theta_1=-30^\circ$ and $\theta_2=30^\circ$, quantum mechanics predicts that the coincidence rate is $\cos^2(60^\circ )=0.25$. That is, the mismatch rate is 75\%, which is not less than the 50\% predicted by the common sense argument.\cite{maxviolation}

Since Bell's publication, measurements that test Bell's inequality have been performed many times,\cite{experiments} and the experimental results clearly agree with quantum mechanics, and violate the inequality. We are therefore forced to reexamine the steps that led to our ``common sense'' conclusions.
	
\section{BELL'S ASSUMPTIONS}

Where did common sense go wrong? There are only a few possibilities in our example. We might wonder if there really exists an angle $\alpha$ for which the mismatch is 25\%. However, its existence is experimentally confirmed and the answer agrees with the prediction of quantum mechanics ($\alpha=30^\circ$). One might also question if the rotated case ($\theta_1=-\alpha$ and $\theta_2=0$) really gives the same 25\% mismatch as the original case ($\theta_1=0$ and $\theta_2=\alpha$), but this case is also experimentally verifiable. These two assumptions are beyond reproach; the vulnerability in the argument lies in the last step, where we argue there are two possibilities. 

To go from the $\theta_1=-\alpha,\ \theta_2=0$ case to the $\theta_1=-\alpha,\ \theta_2=\alpha$ case, we must grab hold of the second polarizer and rotate it from angle 0 to angle $\alpha$. In so doing, we implicitly assume that this action does not affect what happens at filter~1, and that the sequence for filter~1 is the same as it would have been without the rotation of filter~2. This assumption is one of locality.\cite{localityassumptions}

Orthodox quantum mechanics violates this assumption via the process of instantaneous, nonlocal collapse. When photon~2 passes through (or is absorbed by) its filter, the wave function collapses simultaneously for both photons. Photon~1 is suddenly forced to match the polarization of photon~2, which messes up the coincidence rates.

We can see how this change in the coincidence comes about by noting that after the collapse, the zero-degree measurements, which are crucial for comparison to the $\theta_1=\alpha$ and $\theta_2=-\alpha$ measurements, no longer agree between photon~1 and photon~2. If we assume, for example, that photon~1 passes through its $30^\circ$ filter and photon~2 passes through its $-30^\circ$ filter, the collapsed state is then $|30^{\circ}\rangle_1|-30^{\circ}\rangle_2$, which can be expressed in terms of $0^\circ$ and $90^\circ$ states as:
\begin{subequations}
\label{eq:3}
\begin{align}
\psi &= |30^{\circ}\rangle_1|-30^{\circ}\rangle_2\\
&= \left(\frac{\sqrt{3}}{2}|0^{\circ}\rangle_1+\frac{1}{2}|90^{\circ}\rangle_1\right) \left(\frac{\sqrt{3}}{2}|0^{\circ}\rangle_2-\frac{1}{2}|90^{\circ}\rangle_2\right)\\
&= \frac{3}{4}|0^{\circ}\rangle_1|0^{\circ}\rangle_2-\frac{1}{4}|90^{\circ}\rangle_1|90^{\circ}\rangle_2+ \frac{\sqrt{3}}{4}|90^{\circ}\rangle_1|0^{\circ}\rangle_2-\frac{\sqrt{3}}{4}|0^{\circ}\rangle_1|90^{\circ}\rangle_2.
\end{align}
\end{subequations}
In this state there is no longer perfect agreement between photons~1 and 2 in the $0^\circ$--$90^\circ$ basis. The third and fourth terms in Eq.~(3c) show possibilities where one photon passes through a $0^\circ$ filter and the other would not. 

Our previous derivation of Bell's inequality hinged on the assumptions that: 
\begin{enumerate}
\item $\theta_1=0$ measurements of photon~1 are 75\% coincident with $\theta_2=-\alpha$ measurements of photon~2;

\item $\theta_2=0$ measurements of photon~2 are 75\% coincident with $\theta_1=\alpha$ measurements of photon~1; and

\item $\theta_1=0$ measurements of photon~1 are 100\% coincident with $\theta_2=0$ measurements of photon~2.
\end{enumerate}
From these three assumptions, we concluded that $\theta_1=\alpha$ measurements of photon~1 are at least 50\% coincident with $\theta_2=-\alpha$ measurements of photon~2. Now the last assumption in that logic has been broken. The $\theta_1=\theta_2=0$ coincidence between the two photons has been reduced.

Most likely, this particular resolution of the EPR paradox would have set Einstein's magnificent hair on end. It violates locality, one of his most cherished principles. In deference to his coiffure, however, we should note that he was right in one sense: causality itself is still respected. There is still no way for an object to go faster than the speed of light, and no way to send a faster-than-light signal. Causality and special relativity remain intact because wave function collapse is an uncontrollable process. We can measure the polarization of a photon after its collapse, but there is no way to control which of the possible polarizations will be chosen during the collapse. Alice, by measuring and collapsing one photon of an EPR entangled photon pair, cannot in any deliberate way affect the result that Bob will get when he measures the other photon. There is no way to use this feature to send a faster-than-light signal.\cite{Wooters82} Sending information at faster-than-light speeds violates causality, but quantum correlations caused by quantum collapse do not. Mercifully, our grandmothers are safe.

This line of thought has led some authors to assume the laws of physics are nonlocal, but this conclusion is hasty, for there is another subtle assumption hidden in the logic. In the last step of Bell's experiment, we assumed not only that measurements are purely local, but we also assumed a sequence still exists that agrees with both the $\theta_1=30^\circ$ and $\theta_2=-30^\circ$ sequences at the 75\% level. It is this comparison with the (now hypothetical) $\theta_1=\theta_2=0$ sequence that allows us to connect the $\theta_1=30^\circ$ sequence with the $\theta_2=-30^\circ$ sequence in Bell's logic.

Presumably, if we had made the $\theta_1=\theta_2=0$ measurements instead of the $\theta_1=30^\circ ,\ \theta_2=-30^\circ$ measurements, we would have found such a sequence. Moreover, had we been allowed to make the $\theta_1=0,\ \theta_2=-30^\circ$ and $\theta_1=30^\circ, \theta_2=0$ measurements, we could have verified (assuming locality) that the $0^\circ$ measurements agree with the $\theta_1=30^\circ$ and $\theta_2=-30^\circ$ measurements at the 75\% level. However, in the absence of making those measurements, can we still assume such a sequence exists? 

This last question seems so trivial it is difficult to imagine negating it. Part of the question is just inductive reasoning. Can we assume that if we were to repeat the measurements we did earlier, we would obtain results that agree with our previous experience, namely that $\theta=0$ and $\theta=±\alpha$ are coincident 75\% of the time on average? Because this assumption is fundamental to predictive science --- experiments must be repeatable and they must be consistent with each other --- we may safely assume it is acceptable. At least if it fails, there are more serious problems than just the resolution of the EPR experiment. 

The other more subtle part of the question is if we can still postulate getting a single specific well-defined sequence for the $\theta=0$ measurements even when we don't actually make the measurements. Certainly, hidden variable theory satisfies this condition. Hidden variable theory implies that the polarizations of the photons are well defined (a realist assumption) and that they are accurately and precisely described by the extra hidden variables of the theory. In fact, if we knew the values of the hidden variables, not only could we imagine making the $\theta=0$ measurements and obtaining a particular sequence, we could calculate precisely what that sequence would be. But this scenario is overkill. It is not necessary to assume hidden variables, determinism, or even realism to derive Bell's inequality. 

Determinism is demonstrably unnecessary in our example because we assumed only that there is a zero degree sequence that agrees statistically with the previously measured coincidence rate. No specific sequence for the hypothetical zero degree measurements was assumed, only that the sequence remains correlated with $\theta_1=\alpha$ and $\theta_2=-\alpha$ measurements at the 75\% level. The actual sequence in a particular string of measurements could still be random. Any theory that produces random specific results (that still agree with the statistical averages) would satisfy the assumption.\cite{Bell71}

Neither is scientific realism required in the logic. It might be tempting to think that our polarization measurements reveal some underlying reality such as the ``actual'' polarization of the photons. However, our example does not make this assumption. There is no requirement for the existence of photons or any other underlying reality. Our example makes assumptions only about the statistical correlation of measurement sequences. We could repeat the entire argument without ever using the word ``photon,'' making reference only to the behavior of the experimental apparatus. Thus, we might have observed that two sequences produced by identical filter settings are 100\% correlated, and that rotating one filter by angle $\alpha$ produces a 25\% mismatch with the sequence it would have measured before rotation. It follows that two such rotations would produce at most a 50\% mismatch. No reference to any kind of underlying reality (in particular no realist assumption) need be used.

The critical assumption in our example (aside from locality) is that we can still postulate a single definite sequence for the case where one or both filters are oriented at $0^\circ$, even when those measurements are not made. We need not obtain any particular sequence, and the sequence need not be knowable beforehand; we need only specify a sequence that is statistically consistent with previous $0^\circ$ measurements. In other words, we postulate a single sequence for $0^\circ$ orientations (which is identical for both photons) that is still coincident with sequences measured at $+30^\circ$ and $-30^\circ$ at the 75\% level.

This assumption, that we are allowed to postulate a single definite result from an individual measurement even when the measurement is not performed, is known as ``counterfactual definiteness.''\cite{CFDdefinitions} The role of counterfactual definiteness in the Bell argument was first identified by Henry Stapp,\cite{Stapp71} although he claimed to circumvent it in other contexts.\cite{Stappetc} To deny counterfactual definiteness means that we are not necessarily allowed to consider a definite result for an experiment that we have not performed. Stapp originally imagined that the way to violate counterfactual definiteness might be through a kind of ``super determinism,''' where the action of the experimenter is already predetermined, and alternative measurements are not allowed (or even imaginable). In practice, this approach is no different from traditional determinism taken to its logical conclusion. However, this view of counterfactual definiteness is too limited. There is a way to violate it without invoking determinism.

The key is to allow more than one possibility for the potential result of a measurement. Orthodox quantum mechanics embraces this notion of multiple possibilities whenever a quantum state is in a superposition. In the absence of measurement (and collapse), there is no single definite potential result. Instead, there are many potential results represented by many components of the superposition. Moreover, the principle of complementarity guarantees that this situation arises frequently. Complementarity describes how certain pairs of measurements, such as position and momentum measurements or orthogonal spin measurements, cannot be made simultaneously. Thus, a definite momentum measurement forces the physical system into an indefinite superposition of distinct position eigenstates. Likewise, the potential result for one polarization measurement becomes indefinite when making another (non-commuting) polarization measurement. In our example, the unrealized $\theta_1=\theta_2=0$ polarization measurements were rendered indefinite (a superposition of four possibilities) by the act of the $\theta_1=30^\circ$ and $\theta_2=-30^\circ$ measurements. This violation of counterfactual definiteness, together with the violation of locality, gave us the collapsed state in Eq.~(3), which demonstrated the reduced coincidence between photons.

For orthodox quantum mechanics, both locality and counterfactual definiteness are violated, leading naturally to a violation of the Bell inequality. However, it can be confusing to unravel the separate effects of each assumption as they pertain to the Bell experiment. It is possible to violate Bell's inequality using either nonlocality or counterfactual indefiniteness alone, and there are examples of each approach. To better understand the role of counterfactual indefiniteness, it is instructive to examine an interpretation of quantum mechanics that relies solely on counterfactual indefiniteness to violate the inequality. One of the most popular of these is the ``many worlds'' interpretation.
	
\section{MANY WORLDS}

In 1957, Hugh Everett conceived of the ``relative state'' interpretation of quantum mechanics,\cite{Everett57} which was later popularized by Bryce DeWitt\cite{DeWitt73} as the ``many-worlds'' interpretation and further developed by others. The many-worlds interpretation differs from orthodox quantum mechanics by assuming that collapse never happens; superpositions persist forever. The observer, instead of collapsing the measured system to a single result, becomes entangled with the system and is herself described by a more inclusive, entangled wave function with multiple components (that is, ``branches''). Nonetheless, to the observer in a particular branch, who is unaware of other states in other branches, it still looks as if each measurement produces a single result. In this way, the many-worlds interpretation eliminates the nonlocal collapse of orthodox quantum mechanics, while retaining the counterfactual indefinitess of superpositions.

Some features of the many-worlds interpretation are controversial. For our purposes, the most pertinent issue is how to incorporate the idea of probability as represented by the Born rule in orthodox quantum mechanics, which is relevant for our discussion of the statistical correlation between photon polarization measurements. Everett claimed to have answered this challenge by defining probability in terms of the sequences of measurement results seen by observers along the different branches. In Everett's view, each sequence of polarization measurements in each branch is assigned a weight to ``make quantitative statements about the relative frequencies of the different possible results of observation -- which are recorded in the memory --- for a typical observer state.''\cite{Everett57} According to Everett, this weight is represented by the same wave function amplitude that occurs in orthodox quantum mechanics. Thus, although all sequences are present in the many-worlds interpretation, even the rare ones, they have the same relative weights as corresponding sequences in orthodox collapse.

This approach to probability in the many-worlds interpretation has been deemed inadequate by several authors, and some have tried to address the perceived inadequacies of Everett's approach.\cite{Schlosshauer07} For instance, Deutsch\cite{Deutsch99} and Wallace\cite{wallace05} have pursued decision-theoretic methods to deduce the many-worlds probabilities in terms of the experience of a ``rational decision maker.''\cite{Price08} In contrast, Zurek\cite{Zurek05etc} has developed an objective approach to the many-worlds interpretation probability (that is, not subjective to an observer) based on interaction with the environment as a whole, and making use of certain symmetries of entangled states.

It is not the intent of this paper to review the full debate over the many-worlds interpretation. Here we merely wish to note that consistency with observer experience is possible according to many authors, and that collapse-free interpretations such as the many-worlds interpretation can offer a useful perspective for exploring the logic behind Bell's argument. It is also worth noting that following Bell's original work,\cite{Bell66} there have been a number of Bell-like constraints on theory that do not involve inequalities, but instead make precise predictions for common sense theories.\cite{equalities} If we were to pursue any of these experiments, we might claim the discussion of probability is not directly relevant because there is no reference to statistics in the argument. For pedagogical reasons we prefer to focus on the original Bell test and how the many-worlds interpretation might satisfy it, assuming the notion of statistical correlation makes sense in this context. We claim that the many-worlds interpretation passes the Bell test by violating counterfactual definiteness, while still respecting locality.

First of all, we argue that after eliminating the nonlocal collapse of orthodox quantum mechanics, the many-worlds interpretation can be formulated as a local theory. In particular, the correlated entangled states used in EPR-Bell experiments can be produced via purely local processes. For instance, two photons with entangled polarizations might be produced from the decay of a parent particle. In this case the entangled state is produced at one location, where the parent decays, and its immediate effects are limited to that one spacetime point. Thereafter, the photons may go their separate ways, and as they separate they carry the correlation to separate locations. It is the original correlation produced at a single location that guarantees measurements will always match in any experiment in any branch where observers compare notes. In this respect the spread of the correlation to distant locations is akin to the delivery of newspapers, where a common story is generated at a central location and disseminated all over the neighborhood. In the many-worlds context, however, different branches (which originally split at a common location) carry different editions of the newspaper.

In the orthodox interpretation nonlocality appears only when we measure the state of one of the entangled photons. Then and only then, does the state of the other distant photon change in a nonlocal fashion as a result of collapse. In contrast, a measurement in the many-worlds interpretation still involves entanglement --- the measurement apparatus entangles with the object under study thereby branching into an entangled superposition --- but it does not involve collapse, and it remains a purely local process. Measurement induces local branching based on the local measurement result, but it does not cause branching at any distant location. As an example of this thinking, the many-worlds explanation of Bell's experiment\cite{MWIlocality} argues that when measurements are made on a pair of photons with $\theta_1=30^\circ$ for filter~1 and $\theta_2=-30^\circ $ for filter~2, the result is a superposition of four terms:
\begin{eqnarray}
\psi &=& E_1(30^\circ,?)E_2(?,-30^\circ)|30^\circ \rangle_1|-30^\circ \rangle_2+E_1(120^\circ,?)E_2(?,60^\circ)|120^\circ \rangle_1|60^\circ \rangle_2 \nonumber\\
&+& E_1(30^\circ,?)E_2(?,60^\circ )|30^\circ \rangle_1|60^\circ \rangle_2+E_1(120^\circ,?)E_2(?,-30^\circ )|120^\circ \rangle_1|-30^\circ \rangle_2, \label{eq:4}
\end{eqnarray}
where $E_1(30^\circ,?)$ indicates that experimenter 1 has identified photon~1 as being polarized at $30^\circ$ (it has passed through a $30^\circ$ filter) but does not yet know the polarization of photon~2, and similarly for the other terms. At first glance it would seem that each experimenter has branched into four versions -- one branching as a result of the local measurement of his/her own photon and another branching as a result of the distant measurement of the other photon. However, this view is mistaken, for only local branching has really occurred. To illustrate the local nature of that branching, Eq.~(4) can be factored into two terms according to:
\begin{eqnarray}
\psi &=& E_1(30^\circ,?)|30^\circ \rangle_1\left(E_2(?,-30^\circ )|-30^\circ \rangle_2+E_2(?,60^\circ )|60^\circ \rangle_2\right) \nonumber\\
&&{}+ E_1(120^\circ,?)|120^\circ \rangle_1\left(E_2(?,-30^\circ )|-30^\circ \rangle_2+E_2(?,60^\circ )|60^\circ \rangle_2\right), \label{eq:5}
\end{eqnarray}
which indicates that experimenter~1 has split into only two distinct branches according to his local measurement of photon~1. He is not yet entangled (or branched) with the result from photon~2, and has no knowledge of it. Similarly, there are only two branches for experimenter~2, corresponding to her local measurement of photon~2. After the two experimenters communicate their results to each other, each experimenter is finally split into four distinct branches corresponding to the four two-photon states: 
\begin{align}
\psi &= E_1(30^\circ,-30^\circ )E_2(30^\circ,-30^\circ )|30^\circ \rangle_1|-30^\circ \rangle_2+E_1(120^\circ,60^\circ )E_2(120^\circ,60^\circ )|120^\circ \rangle_1|60^\circ \rangle_2 \nonumber\\
&{} \quad + E_1(30^\circ,60^\circ )E_2(30^\circ,60^\circ )|30^\circ \rangle_1|60^\circ \rangle_2+E_1(120^\circ,-30^\circ )E_2(120^\circ,-30^\circ )|120^\circ \rangle_1|-30^\circ \rangle_2, \label{eq:6}
\end{align}
but this final splitting occurs only following a chain of local communications at sublight speed. In this context we see that entanglement and branching in the many-worlds interpretation are local, point-like operations.

If we accept that locality is preserved\cite{nonlocalpropagators} in the many-worlds interpretation, how does the many-worlds interpretation violate Bell's inequality and satisfy the experimental test? We claim that violation of counterfactual definiteness is the key. By keeping superposition and eliminating collapse, the many-worlds interpretation avoids postulating a single result from a measurement (whether or not the measurement is actually made). When a polarization measurement is made, both results (photon transmitted and photon absorbed) exist in separate branches. Postulating a single sequence for a series of polarizations (measured or not) is not valid, and counterfactual definiteness is violated. 

By taking this view, the many-worlds interpretation violates counterfactual definiteness while preserving locality and satisfying the Bell experiment. Perhaps the best way of seeing how this interpretation works in the Bell experiment is to look again at the algebra we used for conventional quantum mechanics. When making a measurement with filter~1 at $30^\circ$ and filter~2 at $-30^\circ$, four different results are possible (each photon is either transmitted or absorbed). The many-worlds interpretation posits that each of the four results persists as a separate component of the superposition. Taken together, the four two-photon states in the four branches comprise the twin state from whence they originated. However, when we consider a particular experimental result, we focus on a single branch, and within this branch the two photons are not in the twin state. For instance, for the branch in which photon~1 passes through a $30^\circ$ filter and photon~2 passes through a $-30^\circ$ filter, the accessible two-photon state is the one in Eq.~(3), just as it is for orthodox quantum mechanics. 

Within this branch, the two-photon state is in a superposition of four possibilities for $\theta_1=\theta_2=0$ measurements, as the four terms of Eq.~(3c) demonstrate. The last two of these terms are cases where the result of the $\theta_1=0$ measurement differs from the result of the $\theta_2=0$ measurement. Thus, in this branch there is no longer a single possibility for $\theta_1=0$ and $\theta_2=0$ measurements that agrees for both photons. In the absence of making the $\theta_1=\theta_2=0$ measurements, we are not allowed to assume a single definite result for those measurements, contrary to the principle of counterfactual definiteness. The alternative possibility where the experimenters make $\theta_1=\theta_2=0$ measurements that always yield the same results for photon~1 and photon~2 is a counterfactual hypothesis, which is not valid in this branch.

In the many-worlds interpretation, in the absence of collapse, there are four branches not one, but there is a loss of correlation in each of the four branches, just as there is for the single result of orthodox quantum mechanics. What is true for the branch described by Eq.~(3) is true for each of the other three branches. By choosing to look at the $\theta_1=30^\circ$ and $\theta_2=-30^\circ$ measurements instead of the $\theta_1=\theta_2=0$ measurements, we have invalidated our previous assumption about the 100\% correlation of the (now counterfactual) $\theta_1=0$ and $\theta_2=0$ measurements. From the point of view of the many-worlds interpretation, we can say that the $\theta_2=0$ sequence that would agree with the $\theta_1=30^\circ$ sequence at the 75\% level and the $\theta_1=0$ sequence that would agree with the $\theta_2=-30^\circ$ sequence at the 75\% level cannot occur in the same branch, and therefore should not be expected to agree with each other.

\section{CONCLUSIONS}

We have argued that the minimal assumptions needed to derive the Bell inequality are locality and counterfactual definiteness. Because nature violates Bell's inequality, we are forced to admit that nature's laws cannot obey both of these assumptions. 

Furthermore, we should note, as did Bell,\cite{Bell87} that the perfect correlations present in EPR experiments imply that local theories must also be deterministic. This conclusion rests on the idea that a theory with only local interactions precludes construction of the EPR correlation at a distance, and therefore the correlation must originate at a common point and propagate outward. For the perfect correlation to propagate undisturbed to separate locations, the theory must also be deterministic. Thus, taking the results of EPR and Bell experiments together, we conclude either that valid physical theories are nonlocal, or they are local, deterministic, and counterfactually indefinite. 

Hidden-variable theories obey counterfactual definiteness. They say not only that specific results to hypothetical experiments can be assumed, but that the values of the hidden variables describe those specific results. In this way, hidden-variable theories define a single-valued realism. Therefore, the only way in which a hidden-variable theory can satisfy Bell's experiment (and violate the inequality) is through a nonlocal effect. This approach describes, for example, David Bohm's ``ontological'' interpretation of quantum mechanics,\cite{Bohm93} which is a nonlocal hidden-variables theory. 

The EPR ``elements of reality'' can be considered a form of hidden variables, at least for the purposes of deriving Bell's inequality. EPR imagined that each ``element'' corresponds to a single specific trait of a physical system. In particular, they did not imagine multiple possibilities (that is, ``superpositions') like those in the many-worlds interpretation or in orthodox quantum mechanics. Thus, the definition of ``elements of reality'' assumed not just scientific realism, but actually a single-valued realism that guarantees definiteness as well.\cite{EPREOR}

The many-worlds interpretation of quantum mechanics satisfies the Bell test in a profoundly different manner. It allows the laws of physics to remain local, at the cost of violating counterfactual definiteness. One may not like the many-worlds interpretation for several reasons (and this author might agree), but it does provide an example in which locality can survive. The many-worlds interpretation is thus realist (in the sense that superpositions can be regarded as real entities and that every possible measurement result exists in some branch), deterministic (the superpositions evolve according to a deterministic wave equation), and local (involving only point-like interactions), but counterfactually indefinite. In this case the multiple possibilities described by the superposition preclude a single definite possibility, and thus provide the means for violating counterfactual definiteness. The many-worlds interpretation is not only counterfactually indefinite, it is factually indefinite as well. Even when measurements are actually performed, many different results can exist in a multitude of different branches. A single definite result is not guaranteed. 
	
The orthodox interpretation violates Bell's inequality by violating both counterfactual definiteness (via superposition and complementarity) and locality (via collapse). It therefore encompasses two methods for getting around Bell's constraints. Incidentally, this interpretation is also non-deterministic and non-realist (the state of a physical system is not necessarily defined until it is collapsed by measurement), but these features are not critical in Bell's analysis.
	
Since Bell's original work, a number of other ``common sense'' constraints on physical theory have been derived,\cite{beyondBell} based on similar thought experiments. However, these derivations rest on assumptions at least as strong as what is in Bell's argument, and their violation doesn't seem to constrain physical reality any further.\cite{Bellcaveat} Some researchers are now focused on discovering new constraints based on assumptions that are distinct from Bell's assumptions. Leggett\cite{Leggett03} has made recent progress in this area by studying a class of nonlocal hidden-variable theories, called ``crypto-nonlocal'' theories, which make some experimental predictions that differ from those of quantum mechanics even though they agree on the Bell test. Gr\"oblacher et al.\cite{Groblacher07} have implemented a version of Leggett's theorem in a real experiment and have ruled out this class of nonlocal hidden-variable theories. In a different approach, Tegmark has written about an experiment he calls ``quantum suicide,'' which might be used to distinguish between the many-worlds interpretation and other interpretations in which measurements produce single outcomes.\cite{Tegmark98} 
	
It appears we have not yet reached the end of research in this field. Researchers are still developing testable theorems about the nature of reality, allowing us to introduce science into what was once the realm of pure philosophy. On this note, we can look forward to a new generation of debate about physical reality that follows in the footsteps of John Bell.

\begin{acknowledgments}
I am grateful to Prof.\ Barry Holstein, Prof.\ Robert Krotkov, Prof.\ Franck Lalo\"{e} and Prof.\ William J.\ Mullin for their thoughtful comments. I would also like to thank the many anonymous reviewers who put in a great deal of time reading (and arguing over) this manuscript.
\end{acknowledgments}


\begin{thebibliography}{99}

\bibitem{Bell66} John S. Bell, ``On the problem of hidden variables in quantum mechanics," Rev. Mod. Phys. {\bf38}, 447--452 (1966).

\bibitem{Einstein35} A. Einstein, B. Podolsky, and N. Rosen, ``Can quantum-mechanical description of physical reality be considered complete?,'' Phys. Rev. {\bf47}, 777--780 (1935).

\bibitem{EPRexamples} Simple examples of such correlated systems might be two particles produced from the decay of a parent at rest, so that their momenta are opposite, or two decay products from a spin singlet parent, so that their spins are opposite.

\bibitem{Bell65} John S. Bell, ``On the Einstein Podolsky Rosen paradox,'' Physics {\bf1}, 195--200 (1965).

\bibitem{experiments} Stuart J. Freedman and John F. Clauser, ``Experimental test of local hidden-variable theories,'' Phys. Rev. Lett. {\bf 28}, 938--941 (1972); G. Faraci, S. Gutowski, S. Notarrigo, and A. R. Pennisi, ``An experimental test of the EPR paradox,'' Lett. Nuovo Cimento {\bf 9}, 607--611 (1974); L. R. Kasday, J. D Ullman, and C. S. Wu, ``Angular correlation of compton-scattered annihilation photons and hidden variables,'' Nuovo Cimento B {\bf 25}, 633--661 (1975); John F. Clauser, ``Experimental investigation of a polarization correlation anomaly,'' Phys. Rev. Lett. {\bf 36}, 1223--1226 (1976); Edward S. Fry and Randall C. Thompson, ``Experimental test of local hidden-variable theories,'' Phys. Rev. Lett. {\bf37}, 465--468 (1976); M. Bruno, M. D'Agostino, and C. Maroni, ``Measurement of linear polarization of positron annihilation photons,'' Nuovo Cimento B {\bf 40}, 143--152 (1977); 
M. Lamehi-Rachti and W. Mittig, ``Quantum mechanics and hidden variables: A test of Bell's inequality by the measurement of the spin correlation in low-energy proton-proton scattering,'' Phys. Rev. D {\bf 14}, 2543--2555 (1976); Alain Aspect, Philippe Grangier, and G\'erard Roger, ``Experimental tests of realistic local theories via Bell's theorem,'' Phys. Rev. Lett. {\bf 47}, 460--463 (1981); Alain Aspect, Philippe Grangier, and G\'erard Roger, ``Experimental realization of Einstein-Podolsky-Rosen-Bohm Gedankenexperiment: A new violation of Bell's inequalities,'' Phys. Rev. Lett. {\bf 49}, 91--94 (1982); Alain Aspect, Jean Dalibard, and G\'erard Roger, ``Experimental test of Bell's inequalities using time-varying analyzers,'' Phys. Rev. Lett. {\bf 49}, 1804--1807 (1982); Y. H. Shih and C. O. Alley, ``New type of Einstein-Podolsky-Rosen-Bohm experiment using pairs of light quanta produced by optical parametric down conversion,'' Phys. Rev. Lett. {\bf 61}, 2921--2924 (1988); Z. Y. Ou and L. Mandel, ``Violation of Bell's inequality and classical probability in a two-photon correlation experiment,'' Phys. Rev. Lett. {\bf 61}, 50--53 (1988); J. G. Rarity and P. R. Tapster,`` Experimental violation of Bell's inequality based on phase and momentum,'' Phys. Rev. Lett. {\bf 64}, 2495--2498 (1990); Z. Y. Ou, S. F. Pereira, H. J. Kimble, and K. C. Peng, ``Realization of the Einstein-Podolsky-Rosen paradox for continuous variables,'' Phys. Rev. Lett. {\bf 68}, 3663--3666 (1992); P. R. Tapster, J. G. Rarity, and P. C. M. Owens, ``Violation of Bell's inequality over 4 km of optical fiber,'' Phys. Rev. Lett. {\bf 73}, 1923--1926 (1994); Paul G. Kwiat, Klaus Mattle, Harald Weinfurter, and Anton Zeilinger, ``New high-intensity source of polarization-entangled photon pairs,'' Phys. Rev. Lett. {\bf 75}, 4337--4341 (1995); W. Tittel, J. Brendel, H. Zbinden, and N. Gisin, ``Violation of Bell inequalities by photons more than 10 km apart,'' Phys. Rev. Lett. {\bf 81}, 3563--3566 (1998); Gregor Weihs, Thomas Jennewein, Christoph Simon, Harald Weinfurter, and Anton Zeilinger, ``Violation of Bell's inequality under strict Einstein locality conditions,'' Phys. Rev. Lett. {\bf 81}, 5039--5043 (1998); M. A. Rowe, D. Kielpinski, V. Meyer, C. A. Sackett, W. M. Itano, C. Monroe, and D. J. Wineland, ``Experimental violation of a Bell's inequality with efficient detection,'' Nature {\bf 409}, 791--794 (2001).

\bibitem{limitations} Much has been written on the limitations of real experiments that purport to test Bell's theorem. I apologetically ignore this discussion to focus on more central issues. See Philippe Grangier, ``Count them all,'' Nature {\bf 409}, 774--775 (2001) for a description of the two most popular complaints and how experimenters have approached them.

\bibitem{HVrealism} Whenever a hidden variable specifies the outcome of an individual measurement, it is naturally taken to represent some form of underlying reality, following logic similar to that used by EPR to define their ``elements of reality.'' In this sense, hidden variable theories are usually considered realist.

\bibitem{HVdeterminism} Hidden-variable theories are usually imagined to be deterministic in the sense that the hidden variables evolve according to deterministic equations, and therefore could be used to predict experimental results. This idea may be what EPR had in mind when they talked of a ``complete'' physical theory. Strictly, however, a hidden variable theory could be non-deterministic; the hidden variables could evolve randomly (possibly even discontinuously) so that their values at one instant do not specify their values at the next instant. Bell referred to this possibility in J. S. Bell, ``Quantum Mechanical Ideas," Science {\bf 177}, 880--881 (1972). 

\bibitem{localrealists} Many physicists seem to believe that Bell's theorem rests on the assumptions of locality and realism. This perspective is found in such notable works as John F. Clauser and Abner Shimony, ``Bell's theorem: experimental tests and implications,'' Rep. Prog. Phys. {\bf 41}, 1881--1927 (1978); Bernard d'Espagnat, ``The quantum theory and reality," Sci. Am. {\bf 241} (5), 158--181 (1979); Jon P. Jarrett, ``On the physical significance of the locality conditions in the Bell arguments,'' No\^us {\bf 18}, 569--589 (1984); M. Ferrero, T. W. Marshall, and E. Santos, ``Bell's theorem: Local realism versus quantum mechanics,'' Am. J. Phys. {\bf 58}, 683--688 (1990); Raymond Y. Chiao and John C. Garrison, ``Realism or locality: Which should we abandon?,'' Found. Phys. {\bf 29}, 553--560 (1999); Tsubasa Ichikawa, Satoshi Tamura, and Izumi Tsutsui, ``Testing the EPR locality using B-mesons,'' Phys. Lett. A {\bf 373}, 39--44 (2008); Abner Shimony, ``Bell's theorem,'' in {\sl The Stanford Encyclopedia of Philosophy} (Fall 2008 edition), edited by Edward N. Zalta, \url{<plato.stanford.edu/archives/fall2008/entries/bell-theorem/>}. In addition, the following papers from Ref.~\onlinecite{experiments} promote the Bell theorem as a test of local realism: Clauser (1976), Lamehi-Rachti and Mittig, Aspect et al.\ (1981), Aspect et al.\ (1982), Rarity and Tapster, Ou et al.\ (1992), Weihs et al., and Rowe et al.

\bibitem{localists} Some discussions of Bell's theorem focus solely on the locality assumption, although some of these authors may have in mind a different definition of locality than what we employ in this paper (see Ref.~\onlinecite{locality}). For example see Nathan Rosen, ``Bell's theorem and quantum mechanics,'' Am. J. Phys. {\bf 62}, 109--110 (1994); Tim Maudlin, {\sl Quantum Non-locality and Relativity} (Blackwell Publishers, Cambridge, 1995); Malcolm Browne, ``Far apart, 2 particles respond faster than light,'' New York Times, July 22, C1--C2 (1997); Charles Seife, ```Spooky action' passes a relativistic test,'' Science {\bf 287}, 1909--1010 (2000); Detlef D\"urr, Sheldon Goldstein, Roderich Tumulka, and Nino Zangh\'i, ``John Bell and Bell's theorem,'' in {\sl The Encyclopedia of Philosophy} (Macmillan Reference, Detroit, MI, 2006), 2nd ed.; Travis Norsen, ``Bell locality and the nonlocal character of nature," Found. Phys. Lett. {\bf 19}, 633--655 (2006); David Z. Albert and Rivka Galchen, ``Was Einstein wrong?: A quantum threat to special relativity,'' Sci. Am. {\bf 300} (3), 32--39 (2009). In addition, the papers by Tapster et al., Kwiat et al., and Tittel et al.\ in Ref.~\onlinecite{experiments} discuss Bell experiments only as a demonstration of nonlocality. 

\bibitem{CFDers} P. H. Eberhard, ``Bell's theorem without hidden variables,'' Nuovo Cimento B {\bf 38}, 75--80 (1977); Asher Peres, ``Unperformed experiments have no results,'' Am. J. Phys. {\bf 46}, 745--747 (1978); Nick Herbert and Jack Karush, ``Generalization of Bell's theorem,'' Found. Phys. {\bf 8}, 313--317 (1978); Brian Skyrms, ``Counterfactual definiteness and local causation,'' Phil. Sci. {\bf 49}, 43--50 (1982); Michael Redhead, {\sl Incompleteness, Nonlocality, and Realism} (Oxford University Press, New York, 1987); Asher Peres, {\sl Quantum Theory: Concepts and Methods} (Kluwer Academic, New York, NY, 1995), Chap. 6; Frank J. Tipler, ``Does quantum nonlocality exist? Bell's theorem and the many-worlds interpretation,'' \url{<xxx.lanl.gov/abs/quant-ph/0003146>}; Mark A. Rubin, ``Locality in the Everett interpretation of Heisenberg-picture quantum mechanics,'' Found. Phys. lett. {\bf 14}, 301--322 (2001); W. E. De Baere, ``On the consequences of retaining the general validity of locality in physical theory,'' Found. Phys. {\bf 35}, 33--56 (2005).

\bibitem{Stapp71} Henry Pierce Stapp, ``S-matrix interpretation of quantum theory,'' Phys. Rev. D {\bf 3}, 1303--1320 (1971).

\bibitem{Bohm51} David Bohm, {\sl Quantum Theory and Measurement} (Prentice Hall, Englewood Cliffs, NJ, 1951), Sec. 16.

\bibitem{polarizers} In practice, there is a problem with measuring photon polarizations using polarizing filters. Only one of the two possible polarization states passes through the filter and is detectable; the orthogonal polarization is absorbed by the filter. For that reason, real experiments use polarization detectors that give a definite signal for both polarization states, such as birefringent crystals that direct the two polarizations in different directions. In this paper, we stick to polarizing filters only because they are more familiar to most readers, and we will pretend that we can positively identify both polarization states.

\bibitem{twinstates} There are several physical processes that produce twin-state photons: parametric down-conversion in which a single high energy photon is converted into a pair of lower energy correlated photons in a non-linear crystal, certain transitions of atomic states (SPS cascades) that emit two photons as the atom decays to the ground state, and annihilation of spin-zero particle states into two gamma rays. It is easy to verify that all of these sources produce photons that satisfy our description of the twin state.

\bibitem{locality} Our definition of ``locality'' based on point-like local interactions is not the only definition used in the literature. For a review, see P. H. Eberhard, ``Bell's theorem and the different concepts of locality,'' Nuovo Cimento B {\bf 46}, 392--418 (1978). Note that Eberhard's first three ``locality properties'' all implicitly assume counterfactual definiteness in addition to an absence of nonlocal effects (see especially Eberhard's discussion of this issue on p. 402). Therefore each of these definitions can be used by itself to derive Bell's inequality. Bell rederived his inequality using a more general approach than local hidden variables in later work [J. S. Bell, ``The theory of local beables," in J. S. Bell, {\sl Speakable and Unspeakable in Quantum Mechanics}, (Cambridge University Press, Cambridge, 1987)] based solely on a concept of ``local causality.'' However, Bell defined ``local causality'' in terms of single-valued ``beables'' that implicitly obey counterfactual definiteness. This condition is violated by any theory that includes superpositions (such as orthodox quantum mechanics or many-worlds quantum mechanics), in which there are multiple possibilities for a measurement result.

\bibitem{FTL} If a faster-than-light signal is identified in one reference frame, there is guaranteed to be another reference frame in which that signal travels backward in time, thus violating causality.

\bibitem{Bellstates} Bell actually described measurements of spin 1/2 particles, but these measurements are logically equivalent to the photon polarization measurements we use in our example.

\bibitem{refframe} Which measurement is identified as the first measurement depends on the reference frame. Here, we assume we are making measurements in the frame in which the two-photon system is described by Eq.~(1). In any other reference frame, the spin entanglement is partially or completely transformed into momentum entanglement. See Robert M. Gingrich and Christoph Adami, ``Quantum entanglement of moving bodies,'' Phys. Rev. Lett. {\bf 89}, 270402-1--4 (2002).

\bibitem{Herbert85} A very similar example of a Bell experiment is given in Nick Herbert's eloquent book: {\sl Quantum Reality: Beyond the New Physics} (Anchor Books, New York, 1985). However, Herbert uses this example only to discuss locality, and does not offer an analysis of any other assumptions in the logic.

\bibitem{inequality} Our example of Bell's inequality derives from comparison of measurements at $0^\circ$ and $\pm30^\circ$. A more general result can be derived for correlations between any arbitrary three angles. Take $C(\theta_1,\theta_2)$ to represent the comparison between measurements of the two photons at angles $\theta_1$ and $\theta_2$, with $C=-1$ identifying a mismatch (one transmission and one absorption) and $C=1$ identifying a coincidence (both absorbed or both transmitted). We can verify that in general $1+C(\theta_1,\theta_3)C(\theta_1,\theta_2)+C(\theta_2,\theta_3)$. If $C(\theta_1,\theta_2) = C(\theta_2,\theta_3) = 1$ (measurements at the three angles are either all absorbed or all transmitted), then $C(\theta_1,\theta_3) = 1$ as well and both sides of the inequality equal 2. If either $C(\theta_1,\theta_2) = -1$ or $C(\theta_2,\theta_3) = -1$, then the left side of the inequality is at most zero and the right side of the inequality is at least zero. By taking averages over many measurements we have $1+\lb C(\theta_1,\theta_3)\rb \lb C(\theta_1,\theta_2)\rb+\lb C(\theta_2,\theta_3)\rb$. In the special case where $\lb C(\theta_1, \theta_2)\rb = \lb C(\theta_2, \theta_3)\rb = 0.5$ (that is, 25\% mismatch), we get $\lb C(\theta_1, \theta_3)\rb ? 0$ (that is, greater than 50\% coincidence), which agrees with our example. There is a sign difference between this result and Bell's original inequality (see Ref.~\onlinecite{Bell65}) only because his result was derived for singlet state fermions, while our result applies to twin state photons. 

\bibitem{maxviolation} The case $\alpha=30^\circ$ gives the maximum difference between quantum mechanics and the inequality limit from Bell's theorem.

\bibitem{localityassumptions} In fact, there are locality assumptions throughout the argument. Initially, we considered measurements with $\theta_1=\theta_2=0$ to verify perfect coincidence between the two photon measurements. Next we rotated filter~2 to angle $\alpha$ to verify (assuming we did not affect the sequence at filter~1) that the $\theta_2=\alpha$ sequence coincides with the $\theta_1=0$ sequence at the 75\% level. Alternatively, we rotated filter~1 to $\theta_1=-\alpha$ to verify (assuming we did not affect sequence at filter~2) that the $\theta_1=-\alpha$ sequence coincides with the $\theta_2=0$ sequence at the 75\% level. Every step in our example involved an implicit locality assumption.

\bibitem{Wooters82} The case against superluminal signaling also rests on the assumption that Bob's entangled photon cannot be copied. This feature of quantum mechanics, known as the ``no-cloning theorem,'' was proven in W. K. Wootters and W. H. Zurek, ``A single quantum cannot be cloned,'' Nature {\bf 299}, 802--803 (1982).

\bibitem{Bell71} Bell emphasized that determinism was not a critical assumption when he published a more general proof of his inequality based on assumptions of local distributions in hidden-variables. See J. S. Bell, ``Introduction to the hidden-variable question," in {\sl Foundations of Quantum Mechanics}, edited by B. d'Espagnat (Academic Press, New York, 1971), pp. 171--181.

\bibitem{CFDdefinitions} An example of counterfactual reasoning is a statement of the form ``If we had made a certain alternative measurement (rather than the one we did make) we would have obtained such-and-such result." Counter factual definiteness implies that a statement such as the former has a definite truth value (is either true or false).

\bibitem{Stappetc} Stapp claimed to circumvent counterfactual definiteness in Henry P. Stapp, ``Nonlocal character of quantum theory,'' Am. J. Phys. {\bf 65}, 300--304 (1997). However, he was repeatedly challenged on this claim. See for instance: N. David Mermin, ``Nonlocal character of quantum theory?,'' Am. J. Phys. {\bf 66}, 920--924 (1998); W. Unruh, ``Nonlocality, counterfactuals, and quantum mechanics,'' Phys. Rev. A {\bf 59}, 126--130 (1999); Lev Vaidman, ``Time-symmetrized counterfactuals in quantum theory,'' Found. Phys. {\bf 29}, 755--765 (1999); A. Shimony and H. Stein, ``Comment on `Nonlocal character of quantum theory' by Henry P. Stapp,'' Am. J. Phys. {\bf 69}, 848--853 (2001).

\bibitem{Everett57} Hugh Everett III, ``Relative state formulation of quantum mechanics,'' Rev. Mod. Phys. {\bf 29}, 454--462 (1957).

\bibitem{DeWitt73} Bryce S. DeWitt and Neill Graham, {\sl The Many-Worlds Interpretation of Quantum Mechanics} (Princeton University Press, Princeton, 1973).

\bibitem{Schlosshauer07} For a thorough review of the debate see Maximilian Schlosshauer, {\sl Decoherence and the Quantum-to-Classical Transition} (Springer, New York, 2007).

\bibitem{Deutsch99} David Deutsch, ``Quantum theory of probability and decisions,'' Proc. R. Soc. Lond. A {\bf 455}, 3129--3137 (1999).

\bibitem{wallace05} David Wallace, ``Quantum probability from subjective likelihood: Improving on Deutsch's proof of the probability rule,'' \url{<xxx.lanl.gov/abs/quant-ph/0312157v2>}.

\bibitem{Price08} For a recent critique of this approach see Huw Price, ``Decisions, decisions, decisions: Can Savage salvage everettian probability?,'' \url{<xxx.lanl.gov/abs/0802.1390v1>}.

\bibitem{Zurek05etc} W. H. Zurek, ``Probabilities from entanglement, Born's rule from envariance,'' Phys. Rev. A {\bf 71}, 52105-1--29 (2005). For further discussion of this approach see M. Schlosshauer and A. Fine, ``On Zurek's derivation of the Born rule,'' Found. Phys. {\bf 35}, 197--213 (2005), and references therein.

\bibitem{equalities} See for example, Daniel A. Greenberger, Michael A. Horne, Abner Shimony, and Anton Zeilinger, ``Bell's theorem without inequalities,'' Am. J. Phys. {\bf 58}, 1131--1143 (1990); Lucien Hardy, ``Nonlocality for 2 particles without inequalities for almost all entangled states,'' Phys. Rev. Lett. {\bf 71}, 1665--1668 (1993); Daniel M. Greenberger, Michael Horne, and Anton Zeilinger, ``A Bell theorem without inequalities for two particles, using efficient detectors,'' \url{<xxx.lanl.gov/abs/quant-ph/0510201>}.

\bibitem{MWIlocality} Our explanation of the many-worlds interpretation branching in the text follows similar descriptions by Don. N. Page, ``The Einstein-Podolsy-Rosen physical reality is completely described by quantum mechanics,'' Phys. Lett. A {\bf 91}, 57--60 (1982), Michael Clive Price, ``The Everett FAQ,'' \url{<www.hedweb.com/manworld.htm>} and C. Hewitt-Horsman and V. Vedral, ``Entanglement without nonlocality,'' Phys. Rev. A {\bf 76}, 062319-1--8 (2007).

\bibitem{nonlocalpropagators} An important caveat is worth mentioning. We have argued that the measurement process of the many-worlds interpretation, namely the branching into different components of a superposition through quantum entanglement, occurs at particular spacetime points, and therefore represents a local process. Strictly, this argument does not rule out the possibility of including other nonlocal effects in the theory. If, for example, the the many-worlds interpretation framework were applied to a relativistic theory that included spacelike propagators (as found in quantum field theories, for instance) we could argue that the resulting theory contains nonlocal effects even though the macroscopic branching obeys relativistic causality. The point here is that nonlocality is not required to satisfy Bell's experiment. 

\bibitem{Bell87} J. S. Bell, ``Bertlmann's socks and the nature of reality," in J. S. Bell, {\sl Speakable and Unspeakable in Quantum Mechanics}, (Cambridge University Press, Cambridge, 1987).

\bibitem{Bohm93} D. Bohm and B. J. Hiley, {\sl The Undivided Universe: An Ontological Interpretation of Quantum Theory} (Routledge, New York, 1993).

\bibitem{EPREOR} If there were no extra assumption (like counterfactual definiteness) in the definition of EPR's ``elements of reality,'' then the ``elements of reality'' would follow directly from an assumption of locality (and also the experimental fact of the EPR correlations, which are undeniable). These in turn could be used to derive Bell's inequality. Following this reasoning, some scientists insist that Bell's inequality rests only on the assumption of locality, and that counterfactual definiteness, which is implied in the definition of ``elements of reality,'' is inferred rather than assumed. This line of thought neglects to realize that the single-reality assumption is already built into the definition of EPR's ``elements.'' Multi-reality interpretations such as many worlds provide a contrasting viewpoint. 

\bibitem{beyondBell} In addition to Ref.~\onlinecite{equalities} see John F. Clauser, Michael A. Horne, Abner Shimony, and Richard A. Holt, ``Proposed experiment to test local hidden-variable theories,'' Phys. Rev. Lett. {\bf 23}, 880--884 (1969) and John F. Clauser and Michael A. Horne, ``Experimental consequences of objective local theories,'' Phys. Rev. {\bf D10}, 526--535 (1974).

\bibitem{Bellcaveat} It is possible that a nonlocal and/or counterfactually indefinite theory might coincidently satisfy Bell's inequality (just because it is nonlocal and/or counterfactually indefinite doesn't mean it must violate the inequality) while violating one of the other constraints. In this case, a theory that passed the Bell test might still be ruled out. However, I know of no particular theory in this category.

\bibitem{Leggett03} A. J. Leggett, ``Nonlocal hidden-variable theories and quantum mechanics: An incompatibility theorem,'' Found. Phys. {\bf 33}, 1469--1493 (2003).

\bibitem{Groblacher07} Simon Gr\"oblacher, Tomasz Paterek, Rainer Kaltenbaek, Caslav Brukner, Marek Zukowski, Markus Aspelmeyer, and Anton Zeilinger, ``An experimental test of nonlocal realism," Nature {\bf 446}, 871--875 (2007).

\bibitem{Tegmark98} See Max Tegmark, ``The interpretation of quantum mechanics: Many worlds or many words?,'' Fortsch. Phys. {\bf 46}, 855--862 (1998). Unfortunately, this clever approach offers proof of the many-worlds-style reality only to the person that performs the experiment (and even then only to the version of that person who survives the suicide attempt). It offers no proof for the rest of the community, nor consolation to the family members who lost their beloved experimenter. Moreover, if the many-worlds interpretation were false, it offers no proof at all.

\end{thebibliography}
\end{document}